\documentclass[11pt, epsfig]{article} 
\usepackage{graphicx}
  \usepackage{amsthm,amsfonts}
  \usepackage{amsmath}
  \usepackage{color}
\newcommand{\bea}   {\begin{eqnarray}}
\newcommand{\eea}   {\end{eqnarray}}

\def\hf{\frac{1}{2}}
\def\del#1{\partial_{#1}}
\def\nn{\nonumber}
\def\g{ \widehat{ \mathfrak{cga} }_1 }
\def\hg{ \widehat{ \mathfrak{g} }_1 }

\def\SDsum{\supset \hspace{-1em}\hspace{-1pt}+}
\def\binom#1#2{ \left( \begin{array}{c} #1 \\ #2 \end{array} \right)}

\begin{document}
\renewcommand{\thefootnote}{\fnsymbol{footnote}}

\thispagestyle{empty}

\title{Invariant PDEs with Two-dimensional Exotic \\ Centrally Extended Conformal Galilei Symmetry}

\author{N. Aizawa\thanks{{E-mail: {\em aizawa@mi.s.osakafu-u.ac.jp}}},\quad 
Z. Kuznetsova\thanks{{E-mail: {\em zhanna.kuznetsova@ufabc.edu.br}}}
\quad and\quad F.
Toppan\thanks{{E-mail: {\em toppan@cbpf.br}}}
\\
\\
}
\maketitle

\centerline{$^{\ast}$
{\it Department of Mathematics and Information Sciences,}}
{\centerline {\it\quad
Graduate School of Science, Osaka Prefecture University, Nakamozu Campus,}}
{\centerline{\it\quad Sakai, Osaka 599-8531 Japan.}}
\centerline{$^{\dag}$
{\it UFABC, Av. dos Estados 5001, Bangu,}}{\centerline {\it\quad
cep 09210-580, Santo Andr\'e (SP), Brazil.}
\centerline{$^{\ddag}$
{\it CBPF, Rua Dr. Xavier Sigaud 150, Urca,}}{\centerline {\it\quad
cep 22290-180, Rio de Janeiro (RJ), Brazil.}
~\\
\maketitle
\begin{abstract} 
Conformal Galilei Algebras labeled by $d,\ell$ (where $d$ is the number of space dimensions and $\ell$ denotes a spin-${\ell}$ representation w.r.t. the $\mathfrak{sl}(2)$ subalgebra) admit two types of central extensions, the ordinary one (for any $d$ and half-integer $\ell$) and the exotic central extension which only exists for $d=2$ and ${\ell}\in\mathbb{N}$.\par
For both types of central extensions invariant second-order PDEs with continuous spectrum were constructed in \cite{AKS}. It was later proved in \cite{AKT1} that the ordinary central extensions also lead to  oscillator-like PDEs with discrete spectrum.\par
We close in this paper the existing gap, constructing \textcolor{black}{a new class of second-order invariant PDEs for the exotic centrally extended CGAs; they admit a discrete and bounded spectrum when applied to a lowest weight representation}. These PDEs are markedly different with respect to their ordinary counterparts. The ${\ell}=1$ case (which is the prototype of this class of extensions, just like the $\ell=\frac{1}{2}$ Schr\"odinger algebra is the prototype of the ordinary centrally extended CGAs) is analyzed in detail.\par

~\\\end{abstract}
\vfill

\rightline{CBPF-NF-007/15}

\newpage
\section{Introduction}

The ${\mathfrak{cga}}_{d,\ell} $ Conformal Galilei Algebras were introduced in \cite{NO}. They are labeled by two parameters, $d$ and ${\ell}$, with $d$ denoting  the number of space dimensions and $\ell$ indicating the spin-$\ell$ representation which accommodates the Abelian ideal generators with respect to the ${\mathfrak{sl}}(2)$ subalgebra. The ${\mathfrak{cga}}_{d,\ell}$ algebras admit two types of central extensions 
\textcolor{black}{(see \cite{MT} for a simple explanation);} 
we have either the ordinary central extension (present for any $d\in {\mathbb N}$ and half-integer $\ell = \frac{1}{2}+\mathbb{N}_0$), or the exotic central extension (only existing for $d=2$ and $\ell\in {\mathbb{N}}$). The centrally extended algebras are denoted as ${\widehat{\mathfrak{cga}}_{d,\ell}}$.\par
The Conformal Galilei Algebras received a lot of attention in connection with the non-relativistic holography (possible application of CFT/AdS duality to, e.g., condensed matter system, see \cite{Hartn}). Their supersymmetric extensions have also been investigated in several papers (see, e.g., \cite{Susy}).\par
Perhaps the most relevant question addressed in the literature is the construction of dynamical systems invariant under the (centrally extended) Conformal Galilei Algebras.  Several results have been obtained, in the classical case and for the ordinary centrally extended CGAs, in a series of papers. It was shown that free higher derivatives theories are invariant under these algebras \cite{Higher} and that, furthermore, they are connected to Pais-Uhlenbeck oscillators at special frequencies \cite{PU}.   \par
The first results in the quantum case were obtained in \cite{AKS}. Using the technique of determining invariant equations from Verma module representations, in that paper, for both types of central extensions,  second-order invariant PDEs with continuous spectrum were obtained (the presence of a non-vanishing central extension is essential for the existence of the invariant PDE). It was later proved in \cite{AKT1}
and \cite{AKT2} that the ordinary centrally extended CGAs (therefore, $\ell=\frac{1}{2}+\mathbb{N}_0$) possess another class of second-order, Schr\"odinger-type, invariant PDEs with a discrete spectrum. For $\ell\geq\frac{3}{2}$ these new equations describe a system of coupled oscillators with a crypto-hermitian (see \cite{AKT2}) Hamiltonian.
\textcolor{black}{At a given $\ell$, the invariant PDEs close an $\mathfrak{sl}(2)$ algebra, the equations in \cite{AKS} resulting in the (positive or negative) root generators, the new PDEs in \cite{{AKT1,AKT2}} being related to the $\mathfrak{sl}(2)$ Cartan generator.} \par
The derivation in \cite{{AKT1,AKT2}} of the invariant PDEs made use of the  so-called {\em on-shell} condition, associated with a given differential realization of the algebra (this technique is also described in \cite{top} and, under a different name, in \cite{{Nied},{Olv}}).\par
In this paper we fill a gap. \textcolor{black}{We apply the on-shell condition (which is described later) to derive, for the exotic ($d=2$ and $\ell\in\mathbb{N}$) centrally extended CGAs, the invariant PDEs related to the $\mathfrak{sl}(2)$ Cartan generator and whose spectrum, computed in a lowest weight representation of ${\widehat{\mathfrak{cga}}}_{d=2,\ell} $, is bounded and discrete.}\par
The resulting PDEs are markedly different with respect to the invariant PDEs related to the ordinary central extensions. The ${\ell}=1$ case is the prototype of this class of exotic extensions (and derived exotic PDEs), playing the same role
as the Schr\"odinger algebra (corresponding to the $\ell=\frac{1}{2}$ case) for the ordinary centrally extended CGAs.\par
We postpone to the Conclusions the comments about our construction and results. We present here the scheme of the paper. In Section {\bf 2} we introduce a differential realization of ${\widehat {\mathfrak{cga}}}_{{d=2},{\ell=1}} $ and discuss the derived  on-shell conditions. \textcolor{black}{ Section {\bf 3} is devoted to discuss at length the (degree $0$) $\ell =1$ invariant PDE whose bounded and discrete spectrum, applied to a lowest weight representation, is derived from a spectrum generating algebra.} In Section {\bf 4} a contraction limit of the discrete spectrum invariant PDE is discussed. In Section {\bf 5} we extend this construction to other integral values of ${\ell}$. In the Conclusions, finally, we summarize our results.

\section{The $\ell=1$ differential realization and the on-shell conditions}

Following \cite{AKS}, a differential realization of the $ d = 2$, $\ell = 1 $ CGA with exotic central extension 
is given by
\bea
  \overline{z}_+ &=& \del{\tau}, \nn \\
  \overline{z}_0 &=& - ( \tau \del{\tau} + x \del{x} + y \del{y} - 1), \nn \\
  \overline{z}_- &=& -\Big( \tau^2 \del{\tau} + 2\tau ( x \del{x} + y \del{y}) 
             + \frac{2}{\xi} x \del{u} + \frac{2}{\gamma} uy + 2 \tau \Big), \nn \\
  \overline{r} &=& - x\del{x} + y \del{y} - u \del{u}, \nn \\
  \overline{v}_{+1} &=& \gamma \del{x}, \qquad 
  \overline{v}_0 = \gamma \Big( \tau \del{x} + \frac{1}{\xi} \del{u} \Big),
   \quad 
  \overline{v}_{-1} =  \gamma \tau^2 \del{x} + \frac{2\gamma}{\xi} \tau \del{u} + \frac{2}{\xi} y, 
   \nn \\             
  \overline{w}_{+1} &=& \xi \del{y}, \qquad 
  \overline{w}_0 =  \xi \Big( \tau \del{y} + \frac{u}{\gamma} \Big), \qquad 
  \overline{w}_{-1} =  \xi \tau^2 \del{y} + \frac{2\xi}{\gamma} \tau u - \frac{2}{\gamma}x,
   \nn \\
  \overline{\theta} &=& 1.
  \label{freerep}
\eea
These first-order differential operators act on the space of functions in $ \tau, x, y $ and $ u.$ 
Their ${\widehat{\mathfrak{cga}}}_{d=2,\ell=1}$ nonvanishing commutators are
\begin{equation}
   \begin{array}{lcl}
      [\overline{z}_0, \overline{z}_{\pm} ] = \pm  \overline{z}_{\pm}, & \qquad & 
      [\overline{z}_+, \overline{z}_- ] = 2  \overline{z}_0, 
      \\[4pt]
      [\overline{z}_0, \overline{v}_k] = k \overline{v}_k, & & 
      [\overline{z}_0, \overline{w}_k] = k  \overline{w}_k,
      \\[4pt]
      [\overline{z}_+, \overline{v}_k]=(1-k)   \overline{v}_{k+1}, & & 
      [\overline{z}_+, \overline{w}_k]=(1-k)   \overline{w}_{k+1},  
      \\[4pt]
      [\overline{z}_-, \overline{v}_k]=(1+k)   \overline{v}_{k-1}, & & 
      [\overline{z}_-, \overline{w}_k]=(1+k)   \overline{w}_{k-1}, 
      \\[4pt]
      [\overline{r}, \overline{v}_k ] = \overline{v}_k, & & 
      [\overline{r}, \overline{w}_k ] = -\overline{w}_k,    
      \\[4pt]
      [\overline{v}_{+1}, \overline{w}_{-1}] = -2 \overline{\theta}, & &   
      [\overline{v}_{0}, \overline{w}_{0}] =  \overline{\theta}, \qquad
      [\overline{v}_{-1}, \overline{w}_{+1}] = -2 \overline{\theta}.
    \end{array}
    \label{CMFree}
\end{equation}
$ \overline{\theta} $ is the (exotic) central element. The operators (\ref{freerep}) coincide, for $  \gamma = \xi = 1 $, with the ones presented in \cite{AKS}. The more general operators (\ref{freerep}) (depending on the two parameters, with arbitrary non-vanishing values, $\xi,\gamma$ ) are recovered by applying the similarity transformation
$g\mapsto SgS^{-1}$, with $S=e^{\alpha_1x\partial_x+\alpha_2y\partial_y}$  for a suitable choice of $\alpha_{1,2}$.\par
A consistent assignment of the scaling dimensions is given by
\bea
&[\xi]=[\gamma]=0, [{\overline z}_\pm ]=\pm 1, [{\overline z}_0]=[{\overline r}]=0,&\nonumber\\ &[{\overline v}_{-1}] =a, [{\overline v}_0] =a+1,
[{\overline v}_{+1}]=a+2,&\nonumber\\& [{\overline w}_{-1}]=-a-2, [{\overline w}_0]=-a-1, [{\overline w}_{+1}]=-a, &\nonumber\\
&[\tau]=-1, [x]=-(a+2), [y]=a, [u]=-(a+1).&
\eea
The special choice $a=-1$ (which implies $[\tau]= [x]=[y]$ and $[u]=0$) corresponds to the grading induced by 
${\overline z}_0$ as degree operator ($[{\overline z}_0,g]=n_gg$, $n_g$ being the degree).\par
  Three second-order on-shell invariant differential operators $ \overline{\Omega}_k \ (k = 0, \pm 1) $ are 
encountered at degree $ k$. They are 
\bea
  \overline{\Omega}_{+1} &=& \overline{z}_+ + \frac{1}{2 }( \{ \overline{v}_{+1}, \overline{w}_0 \} - \{ \overline{v}_0, \overline{w}_{+1} \})
   = \del{\tau} + \xi u \del{x} - \gamma \del{y} \del{u},
  \nn \\
  \overline{\Omega}_0 &=& \overline{z}_0 - \frac{1}{4 }( \{ \overline{v}_{+1}, \overline{w}_{-1} \} - \{ \overline{v}_{-1}, \overline{w}_{+1} \})
   = -   \tau \overline{\Omega}_{+1},
  \nn \\
  \overline{\Omega}_{-1} &=& \overline{z}_- - \frac{1}{2 }( \{ \overline{v}_0, \overline{w}_{-1} \} - \{ \overline{v}_{-1}, \overline{w}_0 \}) 
  = -\tau^2 \overline{\Omega}_{+1}.
  \label{InvOpFree}
\eea
The ${\widehat {\mathfrak{cga}}}_{{d=2},{\ell=1}} $ on-shell invariant condition (see \cite{AKT1, AKT2}) means that the commutators $[g,{\overline{\Omega}_k}]= f_{g,k}\cdot {\overline \Omega}_k$ are satisfied for any 
$g$ in (\ref{freerep}) and for arbitrary functions $f_{g,k}(\tau,x,y,u)$. The on-shell condition is guaranteed in our case due to the fact that the only non-vanishing commutators involving the $ \overline{\Omega}_k $'s and the generators in (\ref{freerep}) are
\bea
  [\overline{z}_0, \overline{\Omega}_k] &=& k   \overline{\Omega}_k, \nn \\[3pt]
  [\overline{z}_+, \overline{\Omega}_0] &=& - \overline{\Omega}_{+1} = \tau^{-1} \overline{\Omega}_0, \nn \\[3pt]
  [\overline{z}_+, \overline{\Omega}_{-1}] &=& 2  \overline{\Omega}_0 = 2 \tau^{-1}  \overline{\Omega}_{-1},
   \nn \\[3pt]
  [\overline{z}_-, \overline{\Omega}_{+1}] &=& -2\overline{\Omega}_0 = 2 \tau \, \overline{\Omega}_{+1},
   \nn \\[3pt]
  [\overline{z}_-, \overline{\Omega}_0] &=& \overline{\Omega}_{-1} =   \tau \, \overline{\Omega}_0. 
  \label{OnShellFree}
\eea
These relations (\ref{OnShellFree}) hold true for any arbitrary non-vanishing values of $ \gamma $ and $ \xi. $ 

 The three operators $ \overline{\Omega}_k $ close the $ \mathfrak{sl}(2) $ algebra, with $ \overline{\Omega}_0 $ as the Cartan element:
\begin{equation}
  [\overline{\Omega}_0, \overline{\Omega}_{\pm 1} ] = \pm   \overline{\Omega}_{\pm 1}, \qquad 
  [\overline{\Omega}_{+1}, \overline{\Omega}_{-1}] = 2   \overline{\Omega}_0.
  \label{sl2Free}
\end{equation}
The degree $1$  differential operator $\Omega_{+1}$ induces the invariant PDE
\begin{equation}
  \overline{\Omega}_{+1} \Psi(\tau, x, y, u) = 0 \quad \equiv \quad 
  \del{\tau} \Psi = ( \gamma \del{y} \del{u} - \xi u \del{x})  \Psi. 
  \label{FreeInvEq}
\end{equation}
For $\gamma=\xi=1$ this invariant PDE was presented in \cite{AKS}.\par
An interesting observation is that the PDE (\ref{FreeInvEq}) has a larger symmetry than ${\widehat {\mathfrak{cga}}}_{{d=2},{\ell=1}} $. Indeed, an extra operator ${\overline q}$ given by
\begin{equation}
  \overline{q} = x \del{y} + \frac{\xi}{2\gamma} u^2, \label{additionalGen}
\end{equation}
commutes with $ \overline{\Omega}_{\pm 1,0} $. Its nonvanishing commutators with the operators in (\ref{freerep}) are
\begin{equation}
   [\overline{r}, \overline{q}] = -2 \overline{q}, \qquad [\overline{v}_k, \overline{q}] = \frac{\gamma}{\xi} \overline{w}_k. 
   \label{additionalNonVaniComm}
\end{equation}
We denote as ``$ \hg$" the enlarged Lie algebra obtained by adding $ \overline{q} $ to the set of
(\ref{freerep}) generators.

\section{The $\ell=1$ invariant PDE with discrete spectrum}

The second order invariant PDE induced by the degree $0$ Cartan operator entering (\ref{FreeInvEq}) is better  expressed in terms of a different differential realization of the $ \hg $ algebra with non-vanishing commutators given by (\ref{CMFree}) and (\ref{additionalNonVaniComm}).\par
This new differential realization acts on functions of 
$t, x, y $ and $u$. We have
\bea
  z_+ &=& e^{-t} (\del{t} - x \del{x} -   y \del{y} ), \nn \\
  z_0 &=& -\del{t} - 1 , \nn \\
  z_- &=& -e^{t} \Big( \del{t} +  x \del{x} +  y \del{y} 
       + \frac{2}{\xi} x \del{u} + \frac{2}{\gamma} u y +2 \Big), 
   \nn \\
  r &=& -x \del{x} + y \del{y} - u \del{u},
   \nn \\
  v_{+1} &=& \gamma e^{-t} \del{x}, \qquad 
  v_0 = \gamma \Big( \del{x} + \frac{1}{\xi} \del{u} \Big),  \qquad 
  v_{-1} = e^{t} \Big( \gamma \del{x} + \frac{2 \gamma}{\xi} \del{u} + \frac{2}{\xi} y \Big),
   \nn \\
  w_{+1} &=& \xi e^{-t} \del{y}, \qquad 
  w_0 = \xi \Big( \del{y} + \frac{1}{\gamma} u \Big), \qquad 
  w_{-1} = e^{t} \Big( \xi \del{y} + \frac{2 \xi}{\gamma} u - \frac{2}{\gamma} x \Big), 
   \nn \\
  \theta  &=& 1, \qquad q = x \del{y} + \frac{\xi}{2\gamma} u^2.  \label{OscRep}
\eea
In this realization the second-order on-shell invariant operators $ \Omega_{\pm 1,0} $ are
\bea
  \Omega_{+1} &=& z_+ + \frac{1}{2}( \{ v_{+1}, w_0 \} - \{ v_0, w_{+1} \}) = -e^{-t} \Omega_0, 
   \nn \\
  \Omega_0 &=& z_0 - \frac{1}{4}( \{ v_{+1}, w_{-1} \} - \{ v_{-1}, w_{+1} \})
   = -\del{t} +  (x \del{x} + y \del{y}) - \xi u \del{x} + \gamma \del{y}\del{u}, 
   \nn \\
  \Omega_{-1} &=& z_- - \frac{1}{2}( \{ v_0, w_{-1} \} - \{ v_{-1}, w_0 \}) = e^{t} \Omega_0.
  \label{InvOpOsci}
\eea
The on-shell invariant condition is guaranteed by the fact that their only non-vanishing commutators with 
the $ \hg $ generators are 
\bea
 [z_0, \Omega_k] &=& k \Omega_k, \nn \\[3pt]
 [z_+, \Omega_0] &=& -\Omega_{+1} =  e^{-t} \, \Omega_0, \nn \\[3pt]
 [z_+, \Omega_{-1}] &=& 2 \Omega_0 = 2 e^{-t} \, \Omega_{-1}, \nn \\[3pt]
 [z_-, \Omega_{+1}] &=& -2 \Omega_0 = 2 e^{t} \Omega_{+1}, \nn \\[3pt]
 [z_-, \Omega_0] &=& \Omega_{-1} =  e^{t} \, \Omega_0. \label{OnShellOsci}
\eea
The $ \Omega_{\pm 1,0} $ operators close the $\mathfrak{sl}(2)$ Lie algebra (\ref{sl2Free}). \par
The two differential realizations (\ref{freerep}, \ref{additionalGen}) versus (\ref{OscRep}) 
of the algebra $ \hg $, as well as their induced invariant differential operators (\ref{InvOpFree}) versus (\ref{InvOpOsci}), 
are connected via a similarity transformation coupled with a redefinition of the time coordinate. It can be described as follows. Let us denote with $g$ an operator entering (\ref{OscRep}}) or (\ref{InvOpOsci}) and with ${\overline g}$ its  corresponding operator
entering (\ref{freerep}), (\ref{additionalGen}) or (\ref{InvOpFree}). 
For convenience we introduce the operator $X_+$ by setting, for $z_+$ in (\ref{OscRep}),
\begin{equation}
  z_+ = e^{-t} ( \del{t} + X_+), \qquad X_+ = - x \del{x} - y \del{y}. 
  \label{XpDef}
\end{equation}
The connection is realized by the similarity transformation and the change of the time coordinate given by
\begin{equation}
  g \ \mapsto \ \bar{g} = e^{t X_+} g e^{-t X_+}, \qquad t \ \mapsto \ \tau =  e^{t}. 
  \label{Free2Osci} 
\end{equation}
%zzzzzzwww
\subsection{The degree $0$ invariant PDE \textcolor{black}{and its spectrum generating algebra}}

In the (\ref{OscRep},\ref{InvOpOsci}) differential realization, the degree 0 invariant equation reads as follows:
\begin{equation}
  \Omega_0 \Psi(t,x,y,u) = 0 \quad \equiv \quad 
  \del{t} \Psi = \big(  x \del{x} + y \del{y} + \gamma \del{y}\del{u} - \xi u \del{x} \big) \Psi.
  \label{OsciInvEq}
\end{equation} 
One can check that the more general three-parameter equation
\begin{equation}
  \del{t} \Psi = \big( x \del{x} + \omega y \del{y} + \gamma \del{y}\del{u} - \xi u \del{x}  \big) \Psi.
  \label{OciEqGeneral}
\end{equation}
is only invariant under ${\widehat{\mathfrak{cga}}}_{2,1} $ (and $ \hg$) for $ \omega = 1 $ (with $\gamma,\xi\neq 0$). 

 The right hand side of (\ref{OsciInvEq}) is the second-order differential operator
\begin{equation}
 H =  x \del{x} + y \del{y} + \gamma \del{y}\del{u} - \xi u \del{x} 
   = \frac{1}{2}(v_{-1} w_{+1} - w_{-1} v_{+1}).
   \label{OscHamil} 
\end{equation}

Under the coordinate transformation $ t = -i s, $ (\ref{OsciInvEq}) gives the \textcolor{black}{equation}
\begin{equation}
  i \del{s} \Psi = H \Psi, \quad H = x \del{x} + y \del{y} + \gamma \del{y}\del{u} - \xi u \del{x}.
  \label{OsciInvSch}
\end{equation} 
The second-order differential operator $H$ in the r.h.s. {\textcolor{black}{does not depend on the ``time" variable $s$. It}} is written, in terms of ${\widehat {\mathfrak{cga}}}_{{2},{1}} $ generators, as
\begin{equation}
  H = \frac{1}{2}(v_{-1} w_{+1} - w_{-1} v_{+1}).
    \label{OscHamil} 
\end{equation}
Its spectrum is computed in an algebraic way. 
\textcolor{black}{The non-vanishing commutators involving the operators $ v_{\pm 1}, w_{\pm 1}, a_{\pm 1}$  are
\begin{equation}
[v_{+1}, w_{-1}]=  [v_{-1}, w_{+1}] = 2[w_0,v_0]=-2.
\end{equation}
}
The commutators of these operators with  $H$ produce
\begin{equation}
   [H, v_{\pm 1}] = \mp v_{\pm 1}, \qquad [H, w_{\pm 1}] = \mp w_{\pm 1}, \qquad
[H,v_0]=[H,w_0] = 0.
   \label{ComOsiHam}
\end{equation}
\textcolor{black}{The ground state $ \psi_0(x,y,u) $ is defined by 
\begin{equation}
  v_{+1} \psi_0 = w_{+1} \psi_0 = v_{0} \psi_0 = 0.
\end{equation} 
}
The state $\psi_{m,n,k}$ ($m,n,k\in {\mathbb N}_0$), defined by 
\begin{equation}
    \psi_{m,n,k} = v_{-1}^m w_{-1}^n w_{0}^k \psi_0, \label{ExciteDef}
\end{equation}
is an eigenstate of the operator $H$ with \textcolor{black}{non-negative integer eigenvalues:}
\begin{equation}
   H \psi_{m,n,k} =  (m+n) \psi_{m,n,k}. \label{OscEigenValues}
\end{equation} 
Each eigenvalue (\ref{OscEigenValues}) is infinitely degenerate due to the \textcolor{black}{creation of the $w_0$ zero modes identified by the quantum number $k.$ The spectrum so identified is discrete and bounded below. It coincides with the (conveniently normalized) spectrum of two harmonic oscillators of the same frequency.} 
 \par
The vector space spanned by the eigenstates (\ref{ExciteDef}) {\textcolor{black}{coincides with the vector space spanned by}} the $ {\cal P}(x,y,u) $ polynomials in $x, y$ and $u$. {\textcolor{black}{The existence of this discrete spectrum for the given vector space is made possible by the association of the (\ref{OsciInvSch}) PDE with the $\mathfrak{sl}(2)$ Cartan generator.}} \par
{\textcolor{black}{It is beyond the scope of this paper to investigate the spectral problem induced by the differential operator (\ref{OscHamil})
when applied to a conveniently defined Hilbert space.}} {\textcolor{black}{This question naturally depends on the class of functions in $x,y,u$ that one is choosing to consider.}} {\textcolor{black}{For instance, at the special $\gamma=\xi=0$ values, for the $x,y$ variables compactified on a $T^2$ torus (therefore, $x=e^{i\vartheta_1}$, $y=e^{i\vartheta_2}$, with $\relax \vartheta_{1,2}\in [0,2\pi ]$) the operator $H$ coincides with the compactified momenta $p_{\vartheta_1} +p_{\vartheta_2}$; its (conveniently normalized) spectrum is given by $E_{n,m} = n+m$, with $n,m\in \mathbb{Z}$.}}\par
On the other hand the vector space spanned by the (\ref{ExciteDef}) eigenfunctions or the $x^ny^mu^k$ monomials coincides, via Gram-Schmidt procedure, with the vector space spanned by the three-variable Hermite polynomials $H_{n,m,k}(x,y,u)$. The inner product which induces the ${\cal L}^2({\mathbb R}^3)$ Hilbert space is defined via the gaussian measure $\int dxdydu e^{-(x^2+y^2+u^2)}$. Since the operator $H$ is not symmetric with respect to this inner product, contrary to the Hermite polynomials, the eigenfunctions (\ref{ExciteDef}) are not orthogonal. It is easily verified that in the ${\cal L}^2({\mathbb R}^3)$ Hilbert space the spectrum of $H$ is no longer discrete, but continuous. Indeed, for any $\lambda$, the function $y^\lambda$ is normalized with respect to the gaussian measure, provided that $\lambda$ is chosen in the range $Re(\lambda) > -\frac{1}{2}$. \par
Since  $Hy^\lambda=\lambda y^\lambda$, the operator (\ref{OscHamil}), defined on ${\cal L}^2({\mathbb R}^3)$, presents a continuous spectrum.

\section{The $\xi= 0$ limit of the discrete spectrum PDE}

The degree 0 invariant PDE (\ref{OsciInvEq}) possesses two free parameters $\gamma $ and $\xi$. 
 A vanishing $ \gamma $ reduces the equation to a first-order PDE.  On the other hand, the $ \xi \to 0$ 
limit keeps a second-order Partial Differential Equation. 
In this limit the variable $x$ (and $t$) becomes a decoupled degree of freedom in the invariant PDE (\ref{OsciInvEq}). 
As a consequence, the invariant PDE has larger symmetry than $ \hg. $ Indeed, one may find an infinite dimensional symmetry algebra. 
An interesting fact is that the $ \xi = 0 $ invariant PDE has the same {\textcolor{black}{ discrete spectrum as the $ \xi \neq 0 $ equation
when applied to a lowest weight representation}} 
(this is expected from the fact that the spectrum is independent of $ \xi$).  
We show at first that the spectrum is the same for $ \xi \neq 0 $ and $ \xi = 0. $ 
Next, we investigate the symmetry of {\textcolor{black}{the}} $ \xi = 0 $ equation. 
 
 The on-shell invariant operator $\Omega_0 $ (\ref{InvOpOsci})  in the $ \xi \to 0$ limit reduces to 
\begin{equation}
   \Omega = -\del{t} +  x\del{x} + y \del{y} + \gamma \del{y} \del{u}.
   \label{ContInvOp}
\end{equation} 
The invariant PDE
\begin{equation}
  \Omega \Psi(t,x,y,u) = 0 \quad \equiv \quad \del{t} \Psi = H \Psi, \quad 
  H =  x\del{x} + y \del{y} + \gamma \del{y} \del{u}
  \label{ContInvEq}
\end{equation}
possesses a second-order differential operator $H$, expressed as in the $ \xi \neq 0 $ case:
\begin{equation}
   H = \frac{1}{2}(v_{-1} w_{+1} - w_{-1} v_{+1}),
   \label{ContHamil}
\end{equation}
where the operators $ v_{\pm 1} $ and $ w_{\pm 1} $ are the $ \xi \to 0 $ limit of (\ref{OscRep}):
\begin{equation}
  v_{+1} = \gamma e^{-t} \del{x}, \quad 
  v_{-1} = 2 e^{t} (\gamma \del{u} +  y ), 
  \quad 
  w_{+1} = e^{-t} \del{y}, \quad
  w_{-1} = -\frac{2}{\gamma} e^{t}x.
  \label{ContRep}
\end{equation}
The limit is taken after rescaling $ v_{-1}, w_{+1} $ as $ \xi v_{-1}, \xi^{-1} w_{+1}. $ 
\textcolor{black}{
Similarly, the $ \xi \to 0 $ limit of the operators $ \xi v_0, \xi^{-1} w_0 $ gives the new operators
\begin{equation}
  v_0 = \gamma \del{u}, \qquad w_0 = \gamma^{-1} u. 
\end{equation}
}
In this way
the commutation relations among $ v_{\pm 1} $ and $ w_{\pm 1} $, as well as the equations  (\ref{ComOsiHam}), are preserved. 
Therefore,  the spectrum of the operator (\ref{ContHamil}) is also given by (\ref{OscEigenValues}) 
\textcolor{black}{ together with the same vector space as $ \xi \neq 0. $ } 

 One may consider a more general setting for $ \xi = 0. $ Since the $x$ degree of freedom is decoupled from $ y, u, $ one may set 
different frequencies for $x$ and $y:$ 
\begin{equation}
   \Omega = z_0 - \frac{1}{4\omega_1} \{v_{+1}, w_{-1}\} + \frac{1}{4\omega_2} \{v_{-1}, w_{+1}\} 
   = -\del{t} + \omega_1 x\del{x} +  \omega_2 y \del{y} + \gamma \del{y} \del{u},
   \label{ContInvOp2}
\end{equation} 
with
\bea
   v_{+1} &=& \gamma e^{-\omega_1 t} \del{x}, \quad 
  v_{-1} = 2\omega_2 e^{\omega_2 t} (\gamma \del{u} + \omega_2 y ), 
  \quad 
  w_{+1} = e^{-\omega_2 t} \del{y}, \quad
  w_{-1} = -\frac{2\omega_1^2}{\gamma} e^{\omega_1 t}x.
  \nn \\
  z_0 &=& -\del{t} - \frac{\omega_1}{2} - \frac{\omega_2}{2}.
  \label{ContDiffFreq}
\eea
The spectrum of this system is $ \omega_1 m + \omega_2n. $ 

The next step consists in considering the symmetries of the degree 0 differential operator (\ref{ContInvOp2}). 
Its on-shell condition can be solved by straightforward computations. \par
The symmetry generators are given by
\bea
 & & -\del{t} + \omega_1 x \del{x}, \nn \\
 & & e^{\omega_2 t} \Big(-\del{t} + \omega_1 x \del{x} - \omega_2 u \del{u} - \frac{\omega_2^2}{\gamma} uy - \omega_2 \Big), 
 \nn \\
 & & e^{-\omega_2 t} (-\del{t} + \omega_1 x \del{x} + \omega_2 y \del{y}), 
 \nn \\
 & & \kappa^k x \del{x}, \qquad \kappa^k (\gamma \del{y} + \omega_2 u), \qquad \ \;
 \kappa^k x^{-\omega_2/\omega_1} (\gamma \del{u} + \omega_2 y), 
 \nn \\
 & & \kappa^k \del{u}, \qquad\ \, \kappa^k (-y \del{y} + u\del{u}), \qquad 
     \kappa^k x^{\omega_2/\omega_1} \del{y},
     \label{ContGenerators}
\eea
where $ \kappa = e^{\omega_2 t} x^{\omega_2/\omega_1} $ and $ k = 0, \pm 1. $ 
The function $ \kappa $ commutes with the differential operator (\ref{ContInvOp2}). 
An infinite number of operators are constructed with the following combinations of (\ref{ContGenerators}) and $ \kappa: $
\bea
  j_0^{(n)} &=& \kappa^n \Big(-\del{t} + \omega_1 x \del{x} -\hf \omega_2 (-y\del{y}+u\del{u}+1) \Big),
   \nn \\
  j_+^{(n)} &=& -\kappa^n e^{-\omega_2 t} (-\del{t} + \omega_1 x \del{x} + \omega_2 y \del{y}),
   \nn \\ 
  j_-^{(n)} &=& \kappa^n \Big( -\del{t} + \omega_1 x \del{x} - \omega_2 u\del{u} - \frac{\omega_2^2}{\gamma} uy - \omega_2 \Big),
   \nn \\
  r^{(n)} &=&  \kappa^n (-y\del{y}+u\del{u}), \qquad\qquad\quad\;\;   
  \chi^{(n)} = \kappa^n x \del{x},
   \nn \\
  w^{(n)} &=&  \kappa^n \Big(\del{y} + \frac{\omega_2}{\gamma} u \Big), \qquad\qquad\qquad \;
  \rho^{(n)} = \kappa^n x^{\omega_2/\omega_1} \del{y},
   \nn \\
  v^{(n)} &=&  \kappa^n x^{-\omega_2/\omega_1} (\gamma \del{u} + \omega_2 y), \qquad\quad
  u^{(n)} = \gamma \kappa^n  \del{u},
   \nn \\
  \theta^{(n)} &=& \kappa^n, \qquad n \in {\mathbb Z}.
   \label{ContInfGen}
\eea
These infinite number of operators give the on-shell invariance of the PDE defined by (\ref{ContInvOp2}). They
close an infinite dimensional Lie algebra. 
Their non-vanishing commutators with $ \Omega $ are
\begin{equation}
  [\Omega, j_+^{(n)}] = -\omega_2 \kappa^n e^{-\omega_2 t} \Omega, 
  \qquad
  [\Omega, j_-^{(n)}] = -\omega_2 \kappa^n e^{\omega_2 t} \Omega.
  \label{ContOnShell}
\end{equation}
By setting $ \Omega_{\pm} = \mp e^{\mp \omega_2 t} \Omega, $ one sees that the operators $ \Omega, \Omega_{\pm} $ close the 
$ \mathfrak{sl}(2)$ Lie algebra
\begin{equation}
  [\Omega, \Omega_{\pm}] = \pm \omega_2 \Omega_{\pm}, \qquad [\Omega_+, \Omega_-] = 2\omega_2 \Omega.
  \label{ContSl2}
\end{equation}
The non-vanishing commutators for (\ref{ContInfGen}) are given by 
\begin{equation}
  \begin{array}{rclcrcl}
   [j_0^{(n)}, j_{\pm}^{(m)}] &=& \pm \omega_2\, j_{\pm}^{(n+m)}, & \qquad & [j+^{(n)}, j_-^{(m)}] &=& 2\omega_2\, j_0^{(n+m)},
   \\[5pt]
   [\chi^{(n)}, \chi^{(m)}] &=& \displaystyle \frac{\omega_2}{\omega_1}(m-n)\, \chi^{(n+m)}, & & 
   [\chi^{(n)}, j_0^{(m)}] &=& \displaystyle \frac{\omega_2}{\omega_1} m\, j_0^{(n+m)},
   \\[10pt]
   [\chi^{(n)}, j_+^{(m)}] &=& \displaystyle \frac{\omega_2}{\omega_1} m\, j_+^{(n+m)}, & &
   [\chi^{(n)}, j_-^{(m)}] &=& \displaystyle \frac{\omega_2}{\omega_1} m\, j_-^{(n+m)},
   \\[10pt]
   [\chi^{(n)}, r^{(m)}] &=& \displaystyle \frac{\omega_2}{\omega_1} m\,r^{(n+m)}, & & & & 
  \end{array}
  \label{InfDimComm1}\nonumber
\end{equation}
\begin{equation}
  \begin{array}{rclcrcl}
     [j_0^{(n)}, w^{(m)}] &=& - \hf \omega_2 \, w^{(n+m)}, & \qquad & 
     [j_0^{(n)}, \rho^{(m)}] &=&  \hf \omega_2 \, \rho^{(n+m)},   
      \\[5pt]
     [j_0^{(n)}, v^{(m)}] &=& - \hf \omega_2 \, v^{(n+m)},  & &
     [j_0^{(n)}, u^{(m)}] &=&  \hf \omega_2 \, u^{(n+m)},
      \\[5pt]
     [j_+^{(n)}, w^{(m)}] &=& \omega_2 \, \rho^{(n+m-1)}, & & 
     [j_+^{(n)}, v^{(m)}] &=& \omega_2 \, u^{(n+m-1)},
      \\[5pt]
     [j_-^{(n)}, \rho^{(m)}] &=& \omega_2 \, w^{(n+m+1)}, & &
     [j_-^{(n)}, u^{(m)}] &=& \omega_2 \, v^{(n+m+1)}, 
  \end{array}
  \label{InfDimComm2}\nonumber
\end{equation}
\begin{equation}
  \begin{array}{rclcrcl}  
     [r^{(n)}, w^{(m)}] &=& w^{(n+m)}, &  \qquad  & [r^{(n)}, \rho^{(m)}] &=& \rho^{(n+m)}, 
     \\[5pt]
     [r^{(n)}, v^{(m)}] &=& -v^{(n+m)},  & & [r^{(n)}, u^{(m)}] &=& -u^{(n+m)}, 
  \end{array}
  \label{InfDimComm3}\nonumber
\end{equation}  
\begin{equation}
  \begin{array}{rclcrcl}  
    [\chi^{(n)}, w^{(m)}] &=& \displaystyle \frac{\omega_2}{\omega_1} m \, w^{(n+m)}, & \qquad &
    [\chi^{(n)}, \rho^{(m)}] &=& \displaystyle \frac{\omega_2}{\omega_1} (m+1) \, \rho^{(n+m)},
    \\[10pt]
    [\chi^{(n)}, u^{(m)}] &=& \displaystyle \frac{\omega_2}{\omega_1} m \, u^{(n+m)}, & \qquad &
    [\chi^{(n)}, v^{(m)}] &=& \displaystyle \frac{\omega_2}{\omega_1} (m-1) \, v^{(n+m)},
    \\[10pt]
    [\chi^{(n)}, \theta^{(m)}] &=& \displaystyle \frac{\omega_2}{\omega_1} m \, \theta^{(n+m)}, & \qquad &
    [\rho^{(n)}, v^{(m)}] &=& \omega_2 \, \theta^{(n+m)},
    \\[7pt]
    [u^{(n)}, w^{(m)}] &=& \omega_2 \, \theta^{(n+m)}. & & & & 
  \end{array}
  \label{InfDimComm4}
\end{equation}  
One may see from (\ref{InfDimComm1}) that     
$ \mathfrak{h}_1 = \{\; j_0^{(n)}, j_{\pm}^{(n)} \;\}, $   
$ \mathfrak{h}_2 = \{ \; \chi^{(n)} \;\} $ and 
$ \mathfrak{h}_3 = \{ \; r^{(n)} \;\} $ close the loop algebra ${{\widetilde{\mathfrak{sl}}}}(2) $ (i.e. the centerless affine ${ \mathfrak{sl}}(2) $ algebra), the Witt algebra and an abelian subalgebra, respectively. They form a closed subalgebra $ \mathfrak{h} = \mathfrak{h}_2 \SDsum (\mathfrak{h}_1 \oplus \mathfrak{h}_3). $   
The set $ \mathfrak{h}_4 =  \{ \; w^{(n)},  u^{(n)}, v^{(n)}, \rho^{(n)}, \theta^{(n)} \;\} $ forms another subalgebra. The  whole structure of this infinite dimensional Lie algebra is $ \mathfrak{h} \SDsum \mathfrak{h}_4. $.

It is worth pointing out that  $\g $ is not a subalgebra of this infinite dimensional Lie algebra.

\section{Extensions to $\ell>1$}

For any integer $\ell$ a differential realization of the algebra $ \widehat{ \mathfrak{cga} }_{\ell}$, which is an extension of (\ref{freerep}), is given by
\bea
  \overline{z}_+ &=& -\del{\tau}, \nn \\
  \overline{z}_0 &=& -\tau \del{\tau} - \sum_{k=1}^{\ell}(\ell+1-k) (x_k \del{x_k} + y_k \del{y_k}) - \hf \ell(\ell+1),
  \nn \\
  \overline{z}_+ &=& 2\tau \overline{z}_0 + \tau^2 \del{\tau} - \ell I_{\ell+1} u y_{\ell}
       - \sum_{k=1}^{\ell} (2\ell+1-k) x_k \del{x_{k+1}}
  \nn \\
  &-& \sum_{k=1}^{\ell-1} (2\ell+1-k) y_k \del{y_{k+1}},
  \nn \\
  \overline{r} &=& \sum_{k=1}^{\ell}( -x_k \del{x_k} + y_k \del{y_k})  -u \del{u},
  \nn \\
  \overline{v}_n &=& \sum_{k=n}^{\ell} \binom{\ell-n}{\ell-k} \tau^{\ell-k} \del{x_{k+1-n}},
  \quad 0 \leq n \leq \ell
  \nn \\
  \overline{w}_n &=& \sum_{k=n}^{\ell} \binom{\ell-n}{\ell-k} \tau^{\ell-k} \del{y_{k+1-n}},
  \quad 1 \leq n \leq \ell
  \nn \\
  \overline{v}_{-n} &=& \sum_{k=0}^{\ell} \binom{\ell+n}{n+k} \tau^{n+k} \del{x_{\ell+1-k}}
    + \sum_{k=1}^n \binom{\ell+n}{n-k} I_{\ell+k} \tau^{n-k} y_{\ell+1-k},
   \quad 1 \leq n \leq \ell
  \nn \\
  \overline{w}_{-n} &=& \sum_{k=1}^{\ell} \binom{\ell+n}{n+k} \tau^{n+k} \del{y_{\ell+1-k}}
    - \sum_{k=0}^n \binom{\ell+n}{n-k} I_{\ell+k} \tau^{n-k} x_{\ell+1-k},
   \quad 0 \leq n \leq \ell  
  \label{freeRepAny}
\eea
where $ x_{\ell+1} = u $ and $ I_n = (-1)^n (2\ell-n)! n!. $ \par
The free parameters $ \gamma $ and $ \xi $ are here set 
equal to $1$. \par
The generators
$ \overline{z}_0, \overline{z}_{\pm} $ close the $ \mathfrak{sl}(2)$ Lie algebra, while the other non-vanishing commutators are
\begin{equation}
   \begin{array}{lcl}
      [\overline{z}_0, \overline{v}_a] = a \overline{v}_a, & \qquad & [\overline{z}_0, \overline{w}_a] = a \overline{w}_a,
      \\[3pt]
      [\overline{z}_+, \overline{v}_a] = (\ell-a) \overline{v}_{a+1}, 
      & & 
      [\overline{z}_+, \overline{w}_a] = (\ell-a) \overline{w}_{a+1},
      \\[3pt]
      [\overline{z}_-, \overline{v}_a] = (\ell+a) \overline{v}_{a-1}, 
      & & 
      [\overline{z}_-, \overline{w}_a] = (\ell+a) \overline{w}_{a-1},
      \\[3pt]
      [\overline{r}, \overline{v}_a] =  \overline{v}_{a},
      & & 
      [\overline{r}, \overline{w}_a] = - \overline{w}_{a}, 
      \\[3pt]
      [\overline{v}_{\pm n}, \overline{w}_{\mp n}] = -I_{\ell+n},
      \label{CMfreeAny}
   \end{array}
\end{equation}
where $ -\ell \leq a \leq \ell $ and $ 0 \leq n \leq \ell. $ 

The degree 1 on-shell invariant differential operator given in \cite{AKS} is 
\begin{equation}
  \overline{\Omega}_1 = \del{\tau} + \sum_{n=1}^{\ell} n x_{n+1} \del{x_n} + \sum_{n=1}^{\ell-1} n y_{n+1} \del{y_n}
  + \frac{ (-1)^{\ell} }{ \ell! (\ell-1)! } \del{y_{\ell}} \del{u}.
  \label{OnShellD1Any}
\end{equation}
We introduce the creation and  annihilation operators
\begin{equation}
  a_n = \overline{v}_n, \qquad a_n^{\dagger} = - \frac{\overline{w}_{-n}}{I_{\ell+n}},
  \qquad 
  b_n = \frac{ \overline{w}_n }{I_{\ell+n}}, \qquad b_n^{\dagger} = \overline{v}_{-n}, 
  \quad n \geq 0.
  \label{CreAnniAny}
\end{equation}
They satisfy the standard canonical commutation relations
\begin{equation}
  [a_n, a_m^{\dagger}] = [b_n, b_m^{\dagger}] = \delta_{nm},
\end{equation}
while all other commutators vanish. At $n=0$ the bosonic operators are not independent. They are related as 
\begin{equation}
    b_0 = - a_0^{\dagger}, \qquad b_0^{\dagger} = a_0.
\end{equation}
One should note that $a_n$ (resp. $b_n$) and $ a_n^{\dagger}$ (resp. $b_n^{\dagger}$) are not mutually hermitian conjugated. The 
annihilation operators with index $ n > 0 $ consist only of differential operators. This implies that the ground state \textcolor{black}{
$
 \varphi $, satisfying  
\bea 
&a_n \varphi = b_n \varphi = 0,& 
\eea
}  can be chosen to be a constant. 
This remains true after performing the  similarity transformation which connects {\textcolor{black}{it to the degree $0$ operator}}.
\par
One can check, at least for low values of $\ell=1,2,3,4$, that, in terms of the creation and annihilation operators, the on-shell invariant operator (\ref{OnShellD1Any}) may 
be written as follows
\begin{equation}
  \overline{\Omega}_1 = \overline{z}_+ + \sum_{n=0}^{\ell-1} \Big( (\ell-n) a_n^{\dagger} a_{n+1}^{\dagger} 
  - (\ell+n+1) b_n^{\dagger} b_{n+1}^{\dagger} \Big).
  \label{OnShellD1Any2}
\end{equation}
From (\ref{CMfreeAny}) and (\ref{CreAnniAny}) we have
\begin{equation}
  \begin{array}{lcl}
     [\overline{z}_{-}, a_n] = (\ell+n) a_n, & & 1 \leq n \leq \ell,
     \\[3pt]
     [\overline{z}_{-}, a_0] = \ell b_1^{\dagger}, & & 
     \\[3pt]
     [\overline{z}_{-}, a_n^{\dagger}] = -(\ell+n+1) a_{n+1}^{\dagger}, & & 0 \leq n \leq \ell-1,
     \\[3pt]
     [\overline{z}_{-}, a_{\ell}^{\dagger}] = 0, & & 
     \\[3pt]
     [\overline{z}_{-}, b_0] = (\ell+1) a_1^{\dagger}, & & 
     \\[3pt]
     [\overline{z}_{-}, b_n] = -(\ell+n+1) b_{n-1}, & & 1 \leq n \leq \ell,
     \\[3pt]
     [\overline{z}_{-}, b_n^{\dagger}] = (\ell-n) b_{n+1}^{\dagger}, & & 0 \leq n \leq \ell .
  \end{array}
\end{equation}
With this we can compute the Cartan operator ${\overline \Omega}_0$ through
\begin{equation}
  [\overline{z}_-, \overline{\Omega}_1] = -2 \overline{z}_0 - 2\sum_{n=1}^{\ell} n (a_n^{\dagger} a_n + b_n^{\dagger} b_n) - \ell(\ell+1)
  = -2 \overline{\Omega}_0.
\end{equation}
Its associated {\textcolor{black}{operator $H$ can be read from this equation}}. We have
\begin{equation}
 H = \sum_{n=1}^{\ell} n (a_n^{\dagger} a_n + b_n^{\dagger} b_n).
\end{equation}
{\textcolor{black}{Its spectrum, obtained from the lowest weight representation,}} is given by
\begin{equation}
  E = \sum_{j=1}^{\ell} j (n_j+m_j)
\end{equation}
with $ n_j, m_j $ non-negative integers.

\section{Conclusions}

We summarize here the main results of the paper. We proved that the $d=2$ exotic, centrally extended Conformal Galilei Algebras admit second-order invariant PDEs {\textcolor{black}{which, applied to a lowest weight representation, furnish a discrete and bounded spectrum}}.\par
Our construction is based on the recognition that, at a given $\ell$, the invariant PDE obtained in \cite{AKS} belongs to a triplet
of invariant PDEs closing an ${\mathfrak{sl}}(2)$ algebra. The {\textcolor{black}{The degree $0$ PDE, producing the discrete spectrum under the specified conditions, is associated with the ${\mathfrak{sl}}(2)$ Cartan generator}}. The main results (for $\ell=1$) are given in Section {\bf 3}.
A convenient first-order differential realization of the ${\mathfrak{cga}}_{d=2,\ell=1}$ algebra,
leading to a second-order invariant equation {\textcolor{black}{whose r.h.s. does not depend on the ``time" variable, was presented}}.
Due to the existence of zero-mode excitations, see equation (\ref{ComOsiHam}), the spectrum is infinitely degenerated. 
{\textcolor{black}{We pointed out that the operator under consideration does not necessarily present a discrete and bounded spectrum if applied to a Hilbert space whose vector space cannot be identified with a lowest weight representation of the spectrum generating algebra}}.\par
{\textcolor{black}{For $\ell=1$ the degree $0$ invariant PDE depends on two non-vanishing, free parameters $\gamma, \xi$ 
entering (\ref{OsciInvSch})}}.  We proved that in the singular $\xi=0$ limit the PDE (which, contrary to the $\gamma=0$ case, is still a second-order differential operator) possesses an infinite dimensional symmetry algebra.   \par
In the last part of the paper we {\textcolor{black}{extended the construction to the $\ell=2,3,4$ degree $0$ invariant PDEs}}. 
\\ {~}~{~}
\par {\Large{\bf Acknowledgements}}
{}~\par{}~\par 
\textcolor{black}{We thank J. P. Gazeau for useful discussions.}
Z.K. and F.T. are grateful to the Osaka Prefecture University, where this work was elaborated, for hospitality.
F.T. received support from CNPq (PQ Grant No. 306333/2013-9). 
N. A. is supported by the  grants-in-aid from JSPS (Contract No. 26400209).

\end{document}